\documentclass[a4paper,fleqn,usenatbib]{mnras}

\usepackage{newtxtext,newtxmath}

\usepackage[T1]{fontenc}
\usepackage{ae,aecompl}

\usepackage{graphicx}	
\usepackage{amsmath}	

\usepackage{dcolumn}
\usepackage{multirow}
\usepackage{hhline}

\usepackage{bm}
\usepackage{booktabs}
\usepackage{adjustbox}

\newcommand{\Cov}{\mathrm{Cov}} 
\newcommand{\LCDM}{$\Lambda$CDM }
\newcommand{\comment}[1]{{}}
\newcommand{\commentout}[1]{{}}

\title[DE from CMB-S4 Lensing and LSST]{The Physical Origin of Dark Energy Constraints from Rubin Observatory and CMB-S4 Lensing Tomography}

\author[B. Yu et al.]{
Byeonghee Yu,$^{1}$ \thanks{E-mail: bhyu@berkeley.edu (BY)}
Simone Ferraro,$^{2,1}$ \thanks{E-mail: sferraro@lbl.gov (SF)}
Z Robert Knight,$^{3}$
Lloyd Knox,$^{3}$
Blake D. Sherwin$^{4,5}$
\\
$^{1}$Berkeley Center for Cosmological Physics, Department of Physics, University of California, Berkeley, CA 94720, USA\\
$^{2}$Lawrence Berkeley National Laboratory, One Cyclotron Road, Berkeley, CA 94720, USA\\
$^{3}$Physics Department, University of California, Davis, CA 95616, USA\\
$^{4}$Department of Applied Mathematics and Theoretical Physics,
University of Cambridge, Wilberforce Road, Cambridge CB3 OWA, UK\\
$^{5}$Kavli Institute for Cosmology Cambridge, Madingley Road, Cambridge CB3 0HA, UK
}

\date{Accepted XXX. Received YYY; in original form ZZZ}

\pubyear{2021}

\begin{document}
\label{firstpage}
\pagerange{\pageref{firstpage}--\pageref{lastpage}}
\maketitle
 
\begin{abstract}

We seek to clarify the origin of constraints on the dark energy equation of state parameter from CMB lensing tomography, that is the combination of galaxy clustering and the cross-correlation of galaxies with CMB lensing in a number of redshift bins. In particular, we consider the two-point correlation functions which can be formed with a catalog of galaxy locations and photometric redshifts from the Vera C. Rubin Observatory Legacy Survey of Space and Time (LSST) and CMB lensing maps from the CMB-S4 experiment. We focus on the analytic understanding of the origin of the constraints. Dark energy information in these data arises from the influence of three primary relationships: distance as a function of redshift (geometry), the amplitude of the power spectrum as a function of redshift (growth), and the power spectrum as a function of wavenumber (shape). 
We find that the effects from geometry and growth play a significant role and partially cancel each other out, while the shape effect is unimportant. 
We also show that Dark Energy Task Force (DETF) Figure of Merit (FoM) forecasts from the combination of LSST galaxies and CMB-S4 lensing are comparable to the forecasts from cosmic shear in the absence of the CMB lensing map, thus providing an important independent check.
Compared to the forecasts with the LSST galaxies alone, combining CMB lensing and LSST clustering information (together with the primary CMB spectra) increases the FoM by roughly a factor of 3-4 in the optimistic scenario where systematics are fully under control. We caution that achieving these forecasts will likely require a full analysis of higher-order biasing, photometric redshift uncertainties, and stringent control of other systematic limitations, which are outside the scope of this work, whose primary purpose is to elucidate the physical origin of the constraints.

\end{abstract}




\section{Introduction}\label{crosscor_Intro}

Future surveys of the Cosmic Microwave Background (CMB) in intensity and polarization will produce high signal-to-noise ratio (SNR) CMB lensing maps over a large fraction of the sky.  The survey conducted from the Atacama Plateau in Chile by CMB-S4 \citep{abazajian2016cmb} will have significant spatial overlap with the deep and wide photometric galaxy catalogs to come from the Vera C. Rubin Observatory, an optical facility located at Cerro Pach\'on, also in Chile \citep{abell2009lsst}. For the first ten years of operation, the Rubin Observatory will perform the Rubin Observatory Legacy Survey of Space and Time (LSST). 

In this paper we investigate physical origin of the constraints on the dark energy equation of state (EoS) parameter $w$ that can be obtained by cross-correlating redshift-binned galaxy maps and a high SNR CMB lensing map. We choose to not include the forecasts of galaxy weak lensing (WL), in order to find out what can be achieved without WL and to create a complementary probe to those which do include WL. We have previously explored the physical origin neutrino mass constraints in a very similar setup in the companion paper \citep{Yu2018}.

Varying the EoS parameter of dark energy affects both the expansion rate of the Universe and the growth of large-scale structure (LSS), which impacts both the amplitude of the matter power spectrum and the angular position of the Baryon Acoustic Oscillation (BAO) features within it.  The distribution of galaxies within a narrow redshift bin traces the distribution of matter at that redshift, and therefore its map and power spectrum contain information about expansion and growth in the same redshift range. On the other hand, the lensing of the CMB traces the distribution of matter over a wide range of redshifts, combined into a single map and power spectrum.  
Galaxy surveys measure the luminous matter, while lensing is sensitive to the underlying matter distribution, so we expect the cross-correlation between galaxies and lensing to provide a measurement of the relationship between luminous and dark matter \citep{abazajian2016cmb}, crucially breaking the intrinsic degeneracy between the amplitude of fluctuations and galaxy bias.  Our goal is to describe the benefit of combining these two sources of information, particularly in how they together can inform us on the dark energy EoS parameter.

The cross-correlation of redshift-binned maps of galaxy number densities
with CMB lensing is very useful due to the different ways in which galaxy clustering and CMB lensing are dependent on the galaxy bias.  We use the standard definition of the linear galaxy bias as the ratio of the overabundance of galaxies to the overdensity of mass, $b(z) = \delta_g(\textbf{r}) / \delta(\textbf{r})$; $\delta_g(\textbf{r}) = (n(\textbf{r}) - \bar{n})/\bar{n}$, where $n(\textbf{r})$ is the density of galaxies at location $\textbf{r}$ and $\bar{n}$ is its spatial average, and $\delta(\textbf{r}) = (\rho(\textbf{r})-\bar{\rho})/\bar{\rho}$, where $\rho(\textbf{r})$ is the mass density at a location \textbf{r} and $\bar{\rho}$ is its spatial average. We can then determine such linear and scale-independent bias, to within noise limitations, as the ratio between angular power spectra, $b_i \simeq C_l^{g_i g_i}/C_l^{\kappa g_i}$, where $i$ runs over tomographic redshift bins. With improved constraints on galaxy bias at various redshifts, we can break the degeneracy between galaxy bias and the amplitude of the matter power spectrum $P(k,z)$, thereby better constraining the cosmological model parameters \citep{pen2004beating, schmittfull2017parameter}.

\comment{
The cross-correlation
of galaxy clustering with CMB lensing is very useful due to the different ways in which galaxy clustering and CMB lensing are dependent on the galaxy bias.  We use the standard definition of the linear galaxy bias as the ratio between the overabundance of galaxies ($\delta_g(\textbf{r}) = (n(\textbf{r}) - \bar{n})/\bar{n}$, where $n(\textbf{r})$ is the density of galaxies at location $\textbf{r}$ and $\bar{n}$ is its spatial average) to the overdensity of dark matter ($\delta(\textbf{r}) = (\rho(\textbf{r})-\bar{\rho})/\bar{\rho}$, where $\rho(\textbf{r})$ is the density of matter at a location \textbf{r} and $\bar{\rho}$ is its spatial average).  For such a linear and scale-independent bias, $b(z) = \delta_g(\textbf{r}) / \delta(\textbf{r})$, where the only remaining dependence is on redshift, we can exploit the relationships of angular power spectra such as the ratio $b_i \simeq C_l^{g_i g_i}/C_l^{\kappa g_i}$, where $i$ runs over tomographic redshift bins, to forecast constraints on the galaxy bias (to within noise limitations).  With improved constraints on galaxy bias at various redshifts, we can then break the degeneracy between galaxy bias and the amplitude of the matter power spectrum $P(k,z)$, and thereby also better constrain the cosmological parameters of a model (such as $\Lambda$CDM$ + \Sigma m_\nu+ w$) which describes that power spectrum \citep{pen2004beating, schmittfull2017parameter}.  
}

\cite{giannantonio2016cmb} used the high depth and density of the DES survey to construct maps of galaxy number density in several photometric redshift bins. They then cross-correlated these maps with a CMB lensing map inferred from \textit{Planck} and SPT data in order to determine both the galaxy bias and the CMB lensing amplitude, in a process they called ``CMB lensing tomography'' (see for example \cite{Sherwin2012, Bleem2012, PlanckLens13} for  early work and \cite{PlanckLens18,Omori2018b,2020JCAP...05..047K, Marques_2020,Darwish21,Hang21,Kitanidis21, 2021arXiv210503421K, 2021arXiv210512108G, Chen:2021vba} for more recent analyses).


CMB lensing tomography provides us with a means, complementary to tomographic cosmic shear, of reconstructing the mass distribution across the sky in coarse slices in redshift.  Here we study the role, in reaching constraints on dark energy parameters, of not just the amplitude as a function of redshift, but also the shape of the matter power spectrum, and the distance-redshift relation that influences observables that are all seen in projection. 

In this paper we focus on the constraints on dark energy that can come from the CMB lensing tomography enabled by CMB-S4 lensing maps and LSST galaxy clustering.
Current SNRs for the best-measured modes in CMB lensing maps are quite modest. The Planck lensing map \citep{ade2016xv}, has a SNR per mode (on spherical harmonic modes with multipole moment $l$) approximately equal to 1 for $l \simeq 50$, and lower everywhere else. From CMB-S4 we expect SNRs of greater than unity for all modes with $l \lesssim 1000$ and as large as $\simeq$ 40 for the best-measured modes. This increase in CMB lensing precision, together with LSST  galaxy clustering, opens up the possibility of measuring the amplitude of structure to a high precision over a range of redshifts \citep{abazajian2016cmb}.

The roles of "geometry" (the distance-redshift relation) and "growth" (the amplitude of the matter power spectrum as a function of time) have been well-studied in the case of tomographic cosmic shear \citep{abazajian2003neutrino,simpson2005illuminating,zhang2005isolating,knox2006distance,zhan2006tomographic,zhan2008distance,matilla2017geometry,zhan2018cosmology}. Although often described as a probe of growth, distinguishing it from purely geometric probes such as the use of SNeIa as standard candles, these studies clarify that geometry is just as important as growth, if not more so, for constraints on dark energy parameters.

\comment{

CMB lensing and galaxy lensing are similar in that they both involve a distribution of lenses across a range of redshifts, but differ in the number of source planes.  Whereas galaxy lensing has a range of sources distributed in redshift, CMB lensing has only a single source plane: the surface of last scattering.  With more than one source plane, galaxy lensing enables one to do a tomographic analysis, providing enough degrees of freedom to map out both the growth of structure and distance as functions of redshift \citep{knox2006distance}.  Galaxy lensing tomography is particularly useful for determining the distance-redshift relationship and breaking the dark matter density - EoS parameter degeneracy (found when analyzing supernovae) \citep{hannestad2006measuring}.  With only one source plane, all of the layers of structure that multiple source planes can elucidate become blended into one map, and so CMB lensing alone is not able to provide constraints as strong as those found from galaxy lensing alone, except in that the matter which causes CMB lensing extends to higher redshift than does that of the observable galaxy lensing.  However, when CMB lensing is cross-correlated with galaxies grouped into redshift bins, part of the difficulty of having only one source plane is taken away, as part of the lensing can be effectively associated with each bin.  

Additional important differences between CMB lensing and galaxy lensing lie in how well we understand the statistics of the sources, and how well we can accurately analyze them.  For CMB lensing, we know the redshift of the source very well, and also have well-understood theories of the unlensed CMB and the statistics which describe it.  On the other hand, the source distribution of galaxies is much less well characterized, troubled by modeling uncertainties such as the accuracy of dark energy corrections to the nonlinear matter power spectrum, high order corrections to the lensing integral, and the effects of baryonic physics on the nonlinear power spectrum, as well as observational systematic uncertainties such as the difficulty of measuring the shapes of galaxies which are angularly very small, the intrinsic alignment of galaxies, photometric redshift uncertainties, shear calibration errors, and point spread function (PSF) uncertainties \citep{joudaki2012dark}.  


} 

Several forecasts have been done for cosmological parameter sets which include $\Sigma m_\nu$ and $w$, through several combinations of observables that include WL, high-SNR CMB lensing, and galaxy clustering.  Early forecasts which included either WL or CMB lensing either did not include galaxy custering \citep{kaplinghat2003determining,hannestad2006measuring,namikawa2010probing,wu2014guide}, or did not include the cross-correlation between CMB lensing and galaxy clustering  \citep{santos2013neutrinos, 2017MNRAS.470.2100K,2021MNRAS.tmp..589E}. More recently, studies have gone to the opposite extreme.  That is, recent studies have included the cross correlation between CMB lensing and galaxy clustering, as part of a robust and inclusive forecast that also includes the cross-correlations between cosmic shear and galaxy clustering, and between cosmic shear and CMB lensing.  However, these studies did not attempt to forecast the benefits of the CMB lensing - galaxy clustering cross-correlation, without also including WL cross-correlations \citep{joudaki2012dark,mishra2018neutrino,schaan4looking}.
 
Two studies which we follow very closely are those of \cite{schmittfull2017parameter} and the companion to this paper, \cite{Yu2018}, in which we presented forecasts of $\Sigma m_\nu$ that include CMB-S4 lensing, LSST galaxy clustering, and their cross-correlation, but did not forecast the dark energy figure of merit. The forecast of \cite{schmittfull2017parameter} includes a  forecast of $\sigma_8$, the linear theory RMS of the mass distribution on scales of 8 Mpc$/h$, which effectively serves as a proxy for the amplitude of the matter power spectrum.   

\cite{Schaan:2020qox} considered the impact of photometric redshift uncertainties and the potential for self-calibration in a setup similar to ours. That work shows that the fraction of photometric redshift outliers can be constrained by the data itself, and it considers the effect of dynamical Dark Energy and neutrinos, similar to this forecast. \cite{Fang:2021ici} explores constraints from cosmic shear, clustering and CMB lensing, including the effect of baryons and photo-$z$ outliers.  However, the physical origin of the constraints on Dark Energy, the main goal of the present paper, was not explored in these works.





\comment{  LK: I think we can just drop the rest of this. I've left it in here in case we decide to resurrect any of it.

The upcoming measurements of the EoS parameter of dark energy are expected to be best constrained by several sources working together, such as the combination of the complementary probes of weak lensing (WL) and Baryon Acoustic Oscillations (BAO).  The LSST will make these and other types of cosmological measurements \citep{zhan2018cosmology}.


CMB lensing is also a powerful probe of dark energy, and here we forecast the usefulness of combining CMB-S4 lensing observations with LSST galaxy observations, in CMB lensing tomography.  Although there is knowledge to be gained about numerous cosmological parameters by combining this information with WL observations \citep{banerjee2018tests, santos2013neutrinos, schaan4looking, mishra2018neutrino}, we take a different approach and instead compare the precision which can be obtained on dark energy parameters by WL, with what can be achieved by CMB lensing tomography.

As modern cosmology moves further into the realm of high precision measurements, the extent to which the uncertainties on the parameters of cosmological models can be reduced is often limited by some systematic aspect of whatever technique is being used to investigate each parameter.  Fortunately, the application of different techniques to the same questions can often be complementary to each other, in that they may be affected by different systematics.  Multiple techniques are thus quite valuable for their potential to reveal systematic errors. 


In addition to comparing the benefits of CMB lensing vs. galaxy lensing, we also are interested in exploring and understanding why CMB lensing correlated with redshift-binned maps of galaxy counts is able to constrain the EoS parameters in ways that galaxy surveys alone can not do.  In order to facilitate this physical understanding, we have used simplified models of the data which will be coming from future observatories.  In particular, we are limiting the data that we examine to the parts of the matter power spectrum on large scales whose evolution is well approximated by linear perturbation theory, and examining parts of the power spectrum which corrections for nonlinear scale dependence in bias are expected to be small, and also working in the limit of zero photometric error.  In future work it will be important to include more accurate bias modeling as well as photo-z errors.  Particularly for our results with $k_{max}=0.15h$Mpc$^{-1}$, we expect use of more accurate modeling to lead to small changes in our forecasts.

We find through the use of Fisher forecasting that the constraints that we can expect to obtain on  the dark energy EoS parameter, when we also add in information that can be obtained about the primary CMB temperature (T) and E-mode polarization (E) maps via CMB-S4, is comparable to the constraints that we also expect through other methods, such as a combination of LSST galaxy shear with CMB T and E observations.

By measuring the BAO features in the galaxy auto spectra, we can create geometric constraints on dark energy parameters.  By cross-correlating power spectra measured in various redshift bins with a high signal-to-noise ratio CMB lensing map, we can break the degeneracy in the growth of the galaxy power spectrum amplitudes between galaxy bias and matter power spectrum growth, \textbf{and measure the growth of structure as a function of time in those bins.  (REALLY? IS THAT WHAT'S HAPPENING?)}  This measurement can be used to help learn more about the mechanism driving the current epoch of cosmic acceleration, such as a variation in the equation-of-state of the dark energy.  The ability to measure the power spectrum amplitude is very important in characterizing dark energy parameters, unlike the case of the sum of the mass of neutrinos, for which the shape of the power spectrum features, which are available without the cross-correlation, is much more important.

CMB-S4 also has other potential path toward $\sigma_8(z)$, $m_\nu$ and dark energy constraints from SZ cluster detection.  What we present here is complementary.
} 

The remainder of our paper is organized as follows: In Section \ref{sec:modelAndAssumptions}, we present our cosmological model space and assumptions about observables and noise. Consistent with our focus on understanding the physics that leads to the forecasted constraints, our modeling of the data is quite simple, and in particular does not include sources of systematic error. 
In Section \ref{sec:fisherForecasting} we describe our forecasting formalism and in Sections \ref{sec:forecastResults} and \ref{sec:discussion} we present and discuss our results. 
Note that we discuss the physical origin of dark energy constraints in Section~\ref{sec:distanceFromGrowth} and \ref{sec:distanceOrGrowth}.

\section{Model and Assumptions}\label{sec:modelAndAssumptions}

In the companion paper \citep{Yu2018}  we assume the \LCDM\ model extended to include massive neutrinos, and here we further extend it to include time-varying dark energy.  The cross-correlations are computed in spherical harmonic space and take the form of angular power spectra.

We assume a fiducial model with the following parameters: 
$\Omega_b h^2 = 0.02226$, \
$\Omega_c h^2 = 0.1193$, \
$\tau = 0.063$, \
$A_s = 2.130\times10^{-9}$, \
$n_s = 0.9653$, \
$\theta_{MC} = 1.04087\times10^{-2}$, and\ 
$\Sigma m_{\nu} = 0.06$ eV.
The neutrinos in this model are relativistic in the early universe and slow down as the universe expands, becoming non-relativistic at late times.
We add time-varying dark energy by considering the dark energy equation of state parameter $w$ to follow the common parameterization \citep{Chevallier:2000qy, Linder:2002et} $w(a) = w_0 + (1-a)w_a$, where $a$ is the scale factor of the expansion of the universe normalized such that $a=1$ today, while noting that different parameterizations are possible \citep{Jassal:2005qc, Efstathiou:1999tm, Barboza:2008rh, Colgain:2021pmf}.  The fiducial values that we use are $w_0 = -1$, $w_a = 0$.  We also include one galaxy bias parameter for each galaxy bin, as described in section~\ref{sec:galaxyBinning}.


\subsection{Galaxy Binning}\label{sec:galaxyBinning}

\begin{figure}
\includegraphics[width=\columnwidth]{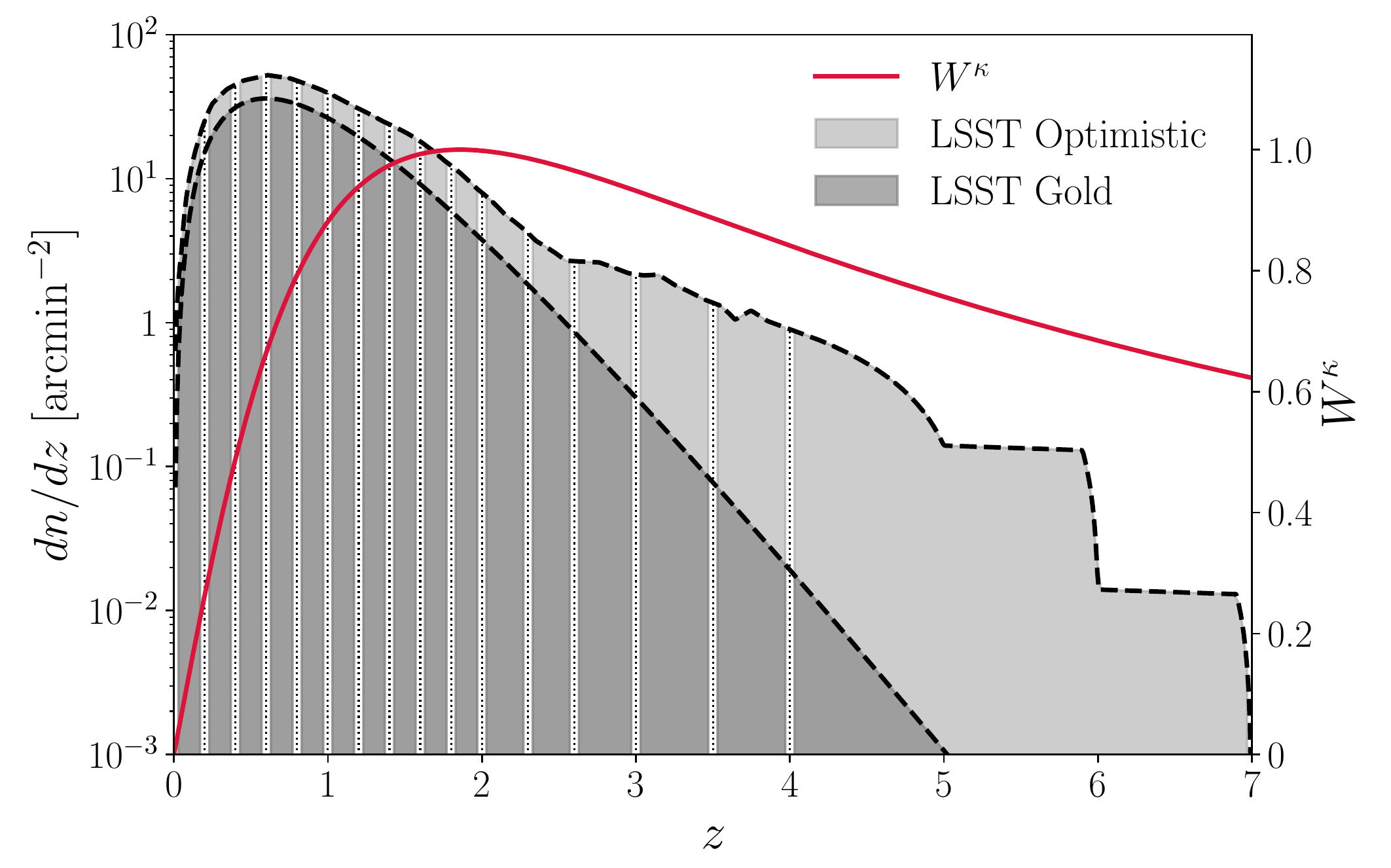}
\caption{The redshift distribution of the CMB lensing convergence kernel (red curve, normalized to a unit maximum) and LSST galaxy samples, both Optimistic (light gray) and Gold (dark gray). We assume 16 tomographic redshift bins in the range $0 < z < 7$, and indicate bin boundaries with vertical lines.  }
\label{fig:dNdz}
\end{figure}

Following \cite{Yu2018}, we employ two different assumptions for LSST galaxy redshift distributions in our analysis, as shown in figure \ref{fig:dNdz}. The first redshift distribution is the $i < 25$ Gold sample \citep{abell2009lsst}, which takes the analytic form $dN(z)/dz \propto (1/2z_0)(z/z_0)^2e^{-z/z_0}$, with $z_0 = 0.3$ and a corresponding galaxy solid angle number density of $\bar n = 40$ arcmin$^{-2}$, and is hereafter referred to as the ``LSST Gold'' sample. The second distribution, which we refer to as the ``LSST Optimistic'', assumes a fainter observable magnitude limit of $i < 27$ with $S/N > 5$ in the $i$ band and includes Lyman break galaxies from redshift dropouts, and this results in the increase the number density of observable galaxies to $\bar n = 66$ arcmin$^{-2}$ \citep{schmittfull2017parameter}. This large number density may not be overly optimistic for our purposes as we do not rely on galaxy shape measurements, but just galaxy locations.


We divide LSST galaxy redshift distributions into 16 non-overlapping tomographic bins with the bin edges of z = [0, 0.2, 0.4, 0.6, 0.8, 1, 1.2, 1.4, 1.6, 1.8, 2, 2.3, 2.6, 3, 3.5, 4, 7].  Defining redshift bins with perfectly sharp edges in an actual galaxy survey, however, is not currently achievable, as the high number of galaxies can only have their redshifts determined photometrically, with an associated photometric redshift error.  Thus our forecasts describe the ideal case in which the photometric redshift (``photo-z'') error has been completely eliminated.  This is a conceptually simpler case than one in which the redshift bins are more realistic and contain the photo-z error, and we leave the forecasting that includes such error to a future study.
 
Likewise, our idealized treatment neglects a number of effects that should be included in more realistic forecasts. These include effects of galactic dust, photometric redshift errors (e.g. \cite{hildebrandt2016kids, Schaan:2020qox}), galaxy-galaxy blending \citep{hartlap2011bias}, and magnification-induced correlations across redshift bins (e.g. \cite{gonzalez2017h}). Also, we neglect any non-Gaussian corrections to the covariance matrix \citep{2013PhRvD..87l3504T, 2017MNRAS.470.2100K}. As stated above, we find that this simplified setting is helpful to elucidate the physical origin of the constraints, and we don't expect our conclusions to change when considering a more realistic forecast.
 

We use galaxies as tracers of matter and assume a single galaxy population. We use a linear galaxy bias model, which \cite{crocce2015galaxy} show to be valid (in the Dark Energy Survey) at least on the scales where the linear growth of structure is a sufficiently accurate approximation. Similarly, we'll restrict our analysis to large scales (defined below), and use the linear matter power spectrum. More sophisticated modeling of non-linearities in matter and bias will be required in a more realistic analysis (see for example \cite{modi17, Krolewski:2021yqy, Kitanidis21, DES:2021zxv}). Following \cite{schmittfull2017parameter}, we use  $b(z) \propto 1+z$ as our fiducial bias evolution.  The exact value of this function should not be important, though higher biases would lead to higher-amplitude galaxy power spectra without increasing the shot noise, and therefore tighter parameter constraints.

For each redshift bin, we define a bias parameter calculated as a weighted average of this galaxy bias function over the redshift range of the bin:

\begin{equation}\label{eqn:bias}
b_i = \frac{1}{\big[ \int  \frac {dN_i(z')}{dz'} dz' \big]} \int \frac {dN_i(z)}{dz} b_i(z) dz,
\end{equation}

\noindent where $b_i(z) = B_i b(z)$. $B_i$ is effectively an amplitude of the bin bias, and $b(z)$ is the redshift dependent galaxy bias function.  The $B_i$ parameters are the ones that we use in our Fisher forecasting, with fiducial values of $B_{i,fid} = 1$.
Similarly, $b_{i,fid} = b_i(B_{i,fid})$,
which leads to $\sigma(B_i) = \sigma(b_i)/b_{i,fid}$ in the Fisher results.

\subsection{Theoretical Power Spectra}\label{sec:powerSpectra}

With the CMB lensing convergence $\kappa$ and a tomographic set of galaxy distribution map, we compute the following 2-point angular power spectra: $C_l^{\kappa \kappa}$, $C_l^{\kappa g_i}$, and $C_l^{g_i g_i}$, where $g_i$ is the galaxy density field in the $i$th tomographic redshift bin. 


The CMB lensing convergence in direction $\mathbf{\hat n}$ can be calculated as a line-of-sight integral over the fractional matter over-density $\delta(\mathbf{r},z)$ at the comoving position $\mathbf{r}$ and redshift $z$:

\begin{equation}\label{eqn:lensingConvergence}
\kappa(\boldsymbol{\hat n}) = \int d\chi W^\kappa(\chi) \delta\big(\chi \boldsymbol{\hat n}, z(\chi)\big),
\end{equation}

\noindent where $\chi$ is the the comoving distance, and the lensing distance kernel is \citep{cooray2000imprint, song2003s, bleem2012measurement}:

\begin{equation}\label{eqn:lensingKernel}
W^\kappa(\chi) = \frac{3}{2} \Omega_m H_0^2 \frac{\chi}{a(\chi)} \frac{\chi_{\text{CMB}}-\chi}{\chi_{\text{CMB}}},
\end{equation}

\noindent where $\Omega_m$ is the matter fraction at the current time, $H_0$ is the current value of the Hubble parameter, $a(\chi)$ is the scale factor at comoving distance $\chi$, and $\chi_{\text{CMB}}$ is the comoving distance of the CMB's surface of last scattering.

For the galaxy density field in the $i$th bin, we can also calculate the following line-of-sight integral:

\begin{equation}\label{eqn:galaxyKernel}
g_i(\boldsymbol{\hat n}) = \int d\chi W^{g_i}(\chi) \delta\big(\chi \boldsymbol{\hat n}, z(\chi)\big),
\end{equation}
where the galaxy distance kernel is \citep{bleem2012measurement}:

\begin{equation}
W^{g_i}(\chi) = \frac{1}{\big[\int dz' \frac{dN_i(z')}{dz'} \big]} \frac{dz}{d\chi} \frac{dN_i(z)}{dz} b_i(\chi).
\end{equation}
Note that we neglect the magnification bias.

Using the Limber approximation \citep{limber1953analysis,kaiser1992weak}, we model the angular power spectrum as:

\begin{equation}
C_l^{\alpha\beta} = \int dz \frac{d\chi}{dz} \frac{1}{\chi^2} W^\alpha(\chi) W^\beta(\chi) P_{\delta_\alpha \delta_\beta}\Big(\frac{l+1/2}{\chi}, z(\chi) \Big)
\end{equation}
where $\alpha, \beta \in (\kappa, g_1, ..., g_N)$, and $P(k, z)$ is the matter power spectrum at wavenumber $k$ and redshift $z$. We use the CDM-baryon density contrast $\delta_{cb}$ for galaxy clustering and the total matter density contrast $\delta_{cb\nu}$ (which includes neutrinos) for lensing.



We use the publicly available \textsc{camb} code \citep{cambWeb,lewis2000efficient} and its Python wrapper \citep{pyCambWeb} in order to calculate $P(k,z)$, as well as the unlensed primary CMB power spectra $C_l^{TT}$, $C_l^{TE}$, and $C_l^{EE}$.  For such calculations, we assume the normal hierarchy, in which the third neutrino mass eigenstate $\nu_3$ is heavier than the other two eigenstates. We use the fluid dark energy model implemented in the python wrapper of \textsc{camb}, available since its major update in version 1.0. 

As in \cite{Yu2018}, we impose the limit on the maximum wavenumber to be included in our analysis (see Table~\ref{tab:k-limits} for the corresponding $l_{\mathrm{max}}$ values of each redshift bin) so that perturbations can be assumed to remain in the linear regime, and therefore uncertainties due to non-linear modeling have negligible effects on our forecasts. All forecasts presented in this work assume the linear matter power spectrum, as we find that the effects of adding non-linear corrections from \textsc{Halofit} to the power spectrum are only negligible with $k_{\mathrm{max}}=0.1$ or 0.2 $h$Mpc$^{-1}$ imposed. 

\newcommand{\ra}[1]{\renewcommand{\arraystretch}{#1}}
\begin{table}
\ra{1.2}
	\centering
	\caption{$l_{\mathrm{max}}$ values corresponding to two $k_{\mathrm{max}}$ limits, 0.1 and 0.2$h$Mpc$^{-1}$, for the left edge of each tomographic redshift bin. ($l_{\mathrm{max}}$ is therefore set to be zero for the first bin.) We assume $l_{\mathrm{min}}=30$ to account for the expected difficulty of attaining low-noise data on large angular scales. }
	\label{tab:k-limits}
\begin{tabular}{ lcccccccc } 
 \toprule
 $k_{\mathrm{max}} \setminus \mathrm{bin}$ & 1 & 2 & 3 & 4 & 5 & 6 & 7 & 8 \\
  \hline
 \phantom{..} 0.1 &  0 & 57 & 108 & 153 & 193 & 229 & 261 & 289 \\
 \phantom{..} 0.2 &  0 & 114 & 216 & 307 & 387 & 459 & 522 & 579 \\
 \hline
 \hline
 $k_{\mathrm{max}} \setminus \mathrm{bin}$ & 9 & 10 & 11 & 12 & 13 & 14 & 15 & 16 \\
 \hline
 \phantom{..} 0.1 & 315 & 338 & 358 & 386 & 411 & 439 & 469 & 495 \\
 \phantom{..} 0.2 & 630 & 676 & 717 & 773 & 822 & 879 & 939 & 991 \\
 \bottomrule
\end{tabular}
\end{table}




\begin{figure}
	\includegraphics[width=\columnwidth]{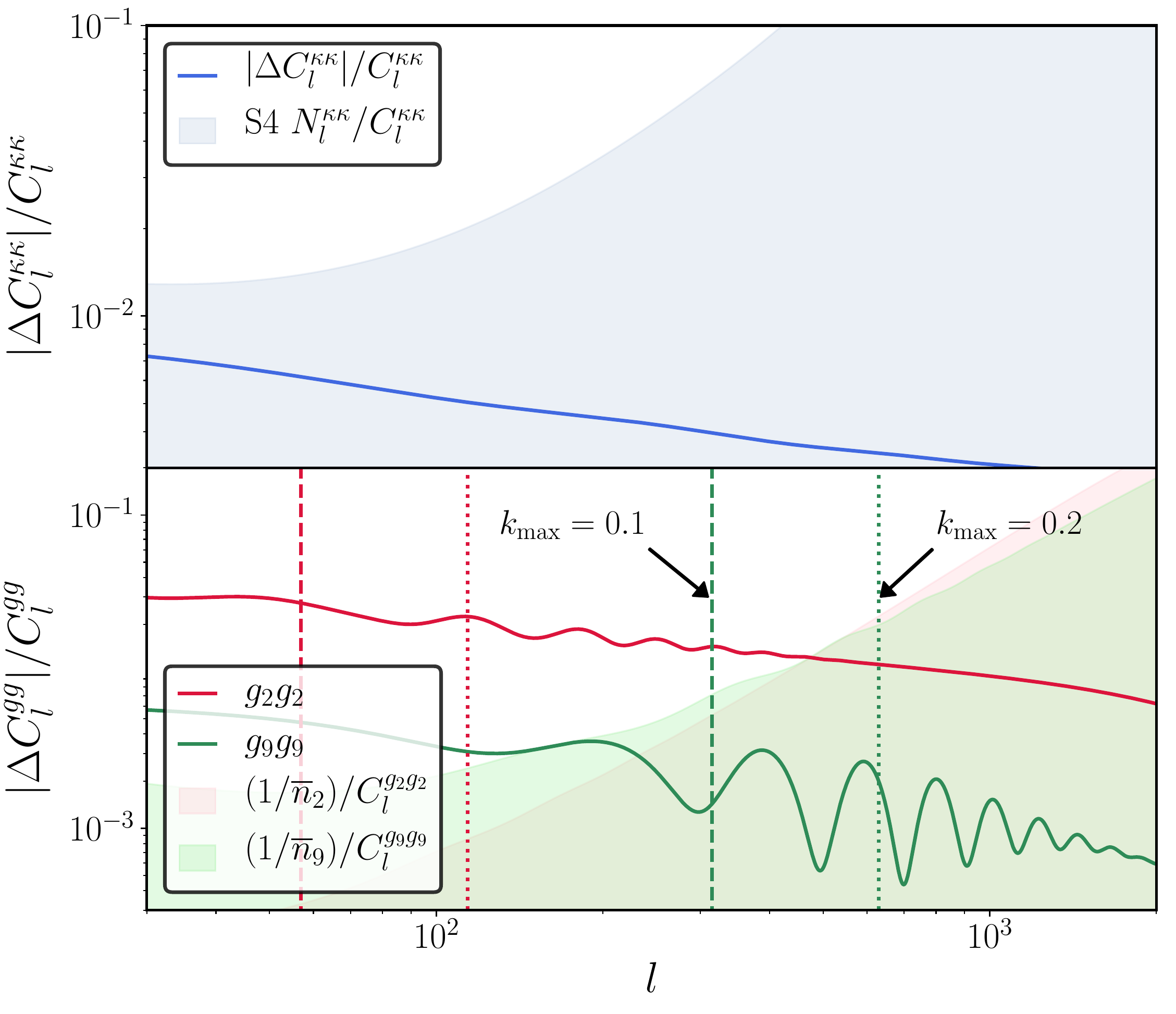}
	\caption{Fractional change of the auto-power spectra $C_l^{\kappa\kappa}$ and $C_l^{gg}$ with respect to $\Delta w_0$: $|\Delta C_l|/C_l = \frac{1}{C_l}|\Delta w_0 \times \partial C_l / \partial w_0|$, where $\Delta w_0 = 0.05$. $\theta_{\mathrm{MC}}$ is held fixed. \textit{Top}: Comparison between the S4 lensing reconstruction noise (blue shaded region) and the changes in the CMB lensing auto-power spectrum with respect to $\Delta w_0$ (blue curve). \textit{Bottom}: The galaxy shot noise and the changes in the galaxy auto-power spectra with respect to $w_0$, in the 2nd (red) and 9th (green) tomographic redshift bin. Also shown are vertical lines indicating the $l_{\mathrm{max}}$ values for each redshift bin, corresponding to the two $k_{\mathrm{max}}$ values indicated in Table \ref{tab:k-limits}.} 
    \label{fig:DeltaCl}
\end{figure}


\subsection{Separating Impacts of Distance, Growth, and Shape}\label{sec:distanceFromGrowth}

The EoS parameter of dark energy affects both the cosmic distance scale and the growth factor.  The relative importance of these two effects on constraining the EoS parameter has been discussed for WL by several groups \citep{abazajian2003neutrino,simpson2005illuminating,zhang2005isolating,knox2006distance,zhan2006tomographic,zhan2008distance,matilla2017geometry}, using various methods.  As was shown by \cite{simpson2005illuminating} and by \cite{matilla2017geometry}, there is a partial cancellation of the geometry and growth effects for WL observables when $w$ is varied.  

\begin{figure}
	\includegraphics[scale=0.36]{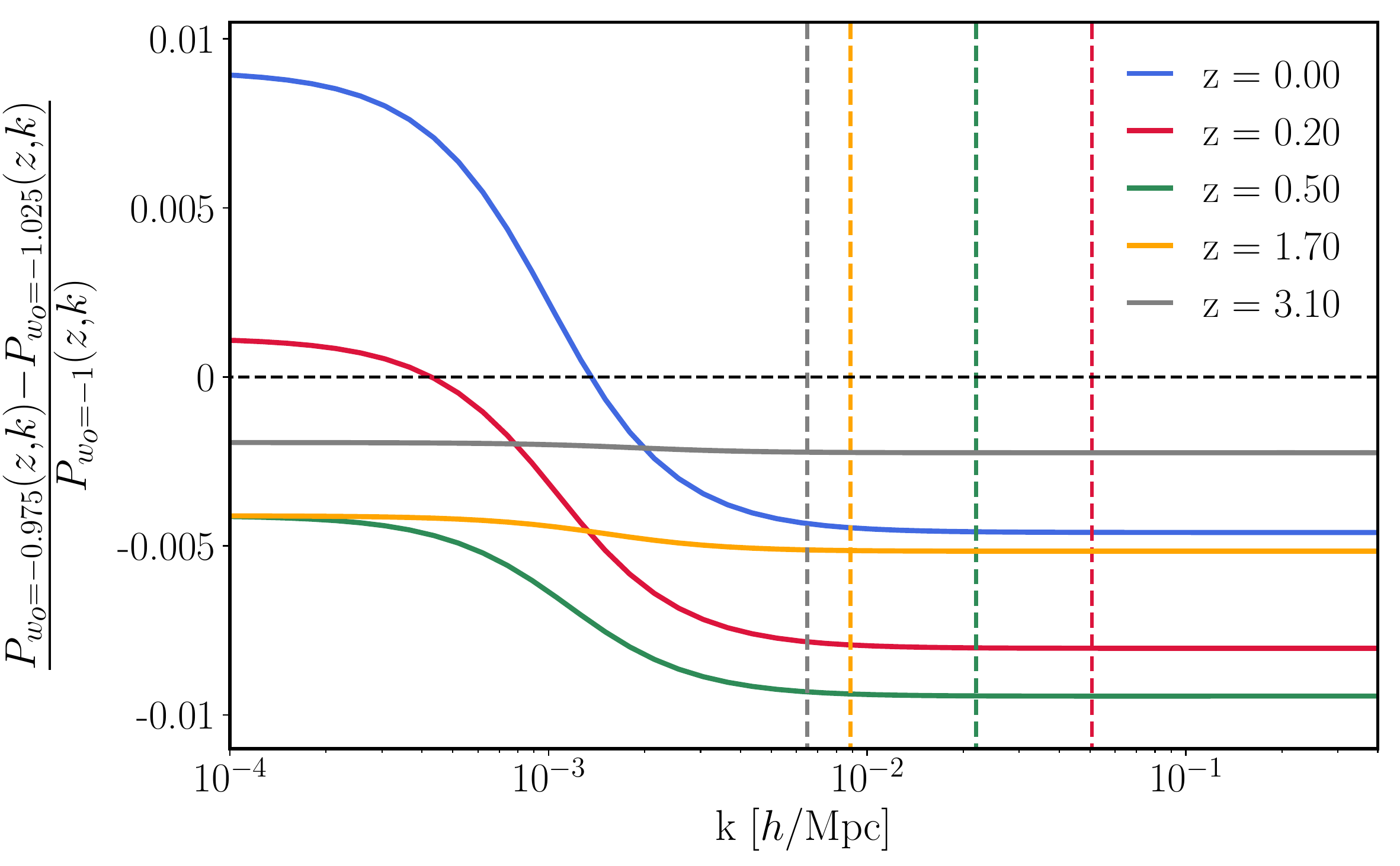}
    \caption{Fractional change of the matter power spectrum $P(z,k)$ with respect to $w_0$ ($\Delta w_0 = 0.05$), $\Delta P(z,k)/P(z,k)$, for five different redshifts within the range of our analysis. 
    $\theta_{\mathrm{MC}}$ is held fixed. To preserve the distance to the last scattering surface, 
    The vertical dashed lines correspond to the low-$l$ cutoff ($l_{\mathrm{min}} = 30$) used in the Fisher forecasts, de-projected to the redshifts (from the right) 0.2, 0.5, 1.7, and 3.1. With the resulting $k$-limits, we remove dark energy perturbations on large scales from the analysis, thereby making the power spectrum shape effects negligible. }
    \label{fig:fractionalChange}
\end{figure}

In Figure~\ref{fig:DeltaCl} and \ref{fig:fractionalChange}, we vary $w_0$ while keeping the angular size of the sound horizon $\theta_{\rm MC}$, $\omega_b$, and $\omega_m$ fixed in order to make minimal changes to primary CMB power spectra. In this scenario, if we increase $w$ from its fiducial value, the dark energy density decreases with time. In order to keep $\theta_{\rm MC}$ fixed, and therefore the angular-diameter distance to last-scattering fixed, we increase the dark energy density at high redshifts. The result is that, compared to the fiducial model, with $w_0$ increased $H(z)$ is decreased at $z \la 0.9$ and gently increased at $z \ga 0.9$, asymptoting to zero increase deep in the dark-matter-dominated regime. One result is that the distances to all redshifts at $z \la 1100$ are increased, asymptoting to zero change deep in the dark-matter dominated regime. Another is that at $z \ga 0.9$, growth is slowed down. The impact on the growth reverses when $H(z)$ starts to become less than in the fiducial model at $z \simeq 0.9$.  Figure~\ref{fig:fractionalChange} indeed shows $\Delta P(k) < 0$ for $\Delta w > 0$, with more power suppression at $z \approx 0.5$ than for any of the other redshift choices shown in the figure. On the other hand, increasing $w$ leads to $D(z) > 0$ at all redshifts (asymptoting to zero at high redshift), thereby increasing the angular power spectrum $C_l$. We find that these two effects partly cancel each other leading to weaker cosmological constraints than if growth and distance were measured separately, as we shall see below. Note that in Figure~\ref{fig:fractionalChange}, $P(k)$ is suppressed at all redshifts, and therefore there is partial cancelation in all redshift bins.

One way of comparing the relative importance of distance and growth on constraining $w$ in a Fisher forecast is to split $w$ into two components: $w_d$, which only affects the distance, and $w_g$, which affects the growth only. Previous work has used this parametrization to distinguish between the two effects and systematic errors \citep{zhang2005isolating}, or to investigate which has more constraining power \citep{simpson2005illuminating, zhan2008distance}.  \cite{zhan2006tomographic} extends this idea of splitting the parameter $w$ to examine constraints on ($w_0, w_a$) for distance-only, growth-only, and full (growth+distance) cases, finding that the growth-only case has the least constraining power; with other cosmological parameters marginalized, the full case has the most constraining power, while fixing them make the distance-only case more powerful.

We instead apply a similar procedure to investigate the relative importance of distance and growth in our forecasts, but we take an additional step of separating the growth effect further into two separate components: the growth of amplitude of the power spectrum and the change in the shape of the power spectrum due to growth.  The reason for such decomposition is the degeneracy between the power spectrum amplitude and the galaxy bias, and we break this degeneracy by incorporating observables which have different dependencies on galaxy bias\footnote{CMB lensing has no galaxy bias dependence whereas galaxy observations do, which allows the combination of the various observables $C_l^{g_i g_i}$, $C_l^{\kappa g_i}$, and $C_l^{\kappa \kappa}$ to distinguish a change in amplitude from the galaxy bias.}.  


Figure~\ref{fig:fractionalChange} shows the fractional change in $P(z,k)$ due with respect to $\Delta w_0$ for five different values of redshift. We only show the plots for $w_0$, but $w_a$ derivatives are very similar in appearance. The absence of scale-invariance seen in this figure is a generic effect associated with $w\neq-1$ dark energy models, as pointed out by e.g. \cite{bean2004probing} and \cite{unnikrishnan2014effect}, who showed that dark energy perturbations appear on very large scales, depend in particular on the sound speed, and are model dependent.

At first glance, Figure~\ref{fig:fractionalChange} looks as if ignoring the shape of $w$ derivatives is effectively flattening the large low-$k$ features in the derivatives. However, these features occur on scales much larger than the maximum angular scale we include in the analysis.  Due to the possibility of large-scale systematics, we also impose a low-$l$ cutoff of $l_{\mathrm{min}} =30$ for each redshift bin. We check that with the resulting $k$-limits, shown as vertical lines in Figure~\ref{fig:fractionalChange}, we remove large perturbative features from the $w$ derivatives. The portion of the power spectrum greater than these cut-off points differs from the shape-less version (identical to the value at $k=0.01h$Mpc$^{-1}$) by only several hundredths of a percent. Due to the smallness of this feature, we can safely neglect it.



When we calculate the partial derivatives of the observables with respect to the $w_0, w_a$ parameters, $P(k,z)$ in each tomographic redshift bin is fixed to the value of $P(k,z_{\mathrm{med}})$, where $z_{\mathrm{med}}$ is the median redshift of each bin, to remove $w(a)$-dependent variations of the power spectrum across the width of each bin. Such variations, degenerate with the evolution of bias across the bin, can make our results artificially sensitive to those changes, especially given that we fix the bias evolution within each bin. We find that overlooking such procedure can lead to artificially rosy forecasts, increasing the constraining power by tens of percent. 

\comment{
Consider the effect of one of our previous assumptions: that we know the way in which galaxy bias varies with redshift and across each redshift bin.  Due to this assumption, and the absence of a similar assumption about the way in which the amplitude of the matter power spectrum varies due to the EoS parameter, a pure amplitude change that varies with redshift due to the EoS parameter is not degenerate with bias. Additionally, since we do not really know that the shape that we have assumed for the way that bias changes across the bin is accurate, we can not rely on this assumption to derive information indicating the way that the shape of $P(k,z)$ changes across the redshift bin.  Because of this, we take out the evolution of the power spectrum over redshift within each bin by assigning the value at the median redshift of each bin to the whole bin.
}




\comment{
\begin{figure}
	\includegraphics[width=\columnwidth]{dPdw_ratio}
    \caption{The ratio of the full power spectrum derivative to the shapeless version, at various redshifts.  The vertical dashed lines correspond to the minimum $l$ value used in the Fisher forecasts, de-projected to the redshifts (from the right) 0.25, 0.5, 1.0, and 2.0.}
    \label{fig:shapeChange}
\end{figure}

\begin{figure}
	\includegraphics[width=\columnwidth]{dPdw_difference}
    \caption{The difference between the derivative of the power spectrum with respect to $w_0$ and that of the shapeless power spectrum, each scaled by the power spectrum, at three different redshifts.  (Since the shapeless power spectrum is defined as being equal to the value at $k=0.1h$Mpc$^{-1}$, this subtraction is the same as subtracting off the shapeless power spectrum.)  Also shown are three vertical lines indicating the low-$l$ cutoffs at those three redshifts (from the right: 0.5, 1.0, 1.5), as in figure \ref{fig:fractionalChange}. The vertical axis is in parts per million, indicating how very tiny these shapes are.}
    \label{fig:shapeChange}
\end{figure}
} 

\section{Fisher Forecasting}\label{sec:fisherForecasting}

We use the Fisher information matrix formalism to forecast constraints on the cosmological parameters of interest \citep{tegmark1997karhunen,bassett2011fisher}.  


\subsection{Observables}\label{sec:observables}

We forecast the constraining power of cross-correlating CMB-S4 lensing with the galaxy clustering tomography observations of Rubin Observatory LSST (similar to \cite{giannantonio2016cmb}), and our observables are auto- ($C_l^{\kappa \kappa}$ and $C_l^{g_i g_i}$) and cross-spectra ($C_l^{\kappa g_i}$) from Section~\ref{sec:powerSpectra}. We do not include $C_l^{g_i g_j}$, cross-spectra of galaxy tomographic bins, nor do we include the cross-spectrum $C_l^{T\kappa}$, which would be nonzero at low $l$ due to the gravitational Integrated Sachs-Wolfe (ISW) effect.

For CMB lensing, we assume a CMB-S4 experiment with the telescope beam of Full-Width-Half-Maximum (FWHM) of $1'$ and a white noise level of $1\mu K'$ for temperature and $1.4\mu K'$ for polarization. We assume $f_{\mathrm{sky}}=0.4$ and set the noise levels $N_l^{TT}$, $N_l^{EE}$ in the primary CMB as a Guassian noise as:

\begin{equation}
N_l^{XX} = s^2 \text{exp} \Big( l(l+1)\frac{\theta_\text{FWHM}^2}{8 \text{log} 2} \Big),
\end{equation}
where $XX$ stands for $TT$ or $EE$, $s$ is the total intensity of instrumental noise in $\mu K$rad, and $\theta_\text{FWHM}^2$ is the FWHM of the beam in radians \citep{wu2014guide}.  For the CMB lensing reconstruction noise, we use the EB quadratic estimator method described in \cite{hu2002mass}, implemented by the \textsc{quicklens} \citep{quicklensWeb} software package.  Following \cite{schmittfull2017parameter}, we rescale the EB noise to approximately match the expected improvement from iterative lens reconstruction for CMB-S4 \citep{hirata2003reconstruction,smith2012delensing}.  

For the LSST, we assume that the survey covers an area on the sky of 18,000 deg$^2$, corresponding to $\approx$ 40\% of the sky, and that it fully overlaps with CMB-S4. The shot noise associated with the galaxy redshift distributions is $1/\bar{n}_i$, where $\bar{n}_i$ is the galaxy number density per redshift bin, calculated per bin from the ratio of the integrated area of $dN_i/dz$ to that of the total $dN(z)/dz$ multiplied by the overall galaxy density $\bar{n}$.




\subsection{Fisher Matrices}\label{sec:fisherMatrices}

Assuming our observables from Section~\ref{sec:observables} are the power spectra of Gaussian random fields, we can compute the covariance matrix as:
\begin{equation}
\Cov(C_l^{\mu_1\nu_1}, C_{l'}^{\mu_2\nu_2} ) =  \frac{ \delta_{ll'}}{(2l+1)f_\text{sky}}  \Big( C_l^{\mu_1\mu_2} C_l^{\nu_1\nu_2} + C_l^{\mu_1\nu_2} C_l^{\nu_1\mu_2} \Big),
\end{equation}
where $(\mu_1,\mu_2,\nu_1,\nu_2) \in \{\kappa, g_1, ..., g_N\}$. We assume that each $C_l$ contains both signal and noise. 

Then, the Fisher matrix is given by:
\begin{align}
F_{ij} = \sum_{\substack{\mu_1, \nu_1, \\ \mu_2, \nu_2}} \sum_{l} \frac{\partial C_{l}^{\mu_1 \nu_1}}{\partial \theta_i} \Big[\Cov(C_l^{\mu_1\nu_1}, C_{l'}^{\mu_2\nu_2} )\Big]^{-1} \frac{\partial C_{l}^{\mu_2 \nu_2}}{\partial \theta_j},
\label{eq:Fisher}
\end{align}
where $\vec{\theta}$ is a set of cosmological model parameters from Section~\ref{sec:modelAndAssumptions}. We can combine the Fisher matrix in equation~\ref{eq:Fisher} with external datesets, such as the primordial CMB and BAO Fisher matrices, if needed and then invert the resulting matrix to determine the marginalized constraints on the parameters of our interest.

\begin{table*}
\ra{1.3}
\centering
\begin{tabular}{l cc c cc cc cc cc cc}
\toprule
  & \multicolumn{2}{c}{$\Lambda$CDM + $m_{\nu}$ free} && \multicolumn{10}{c}{+ $w_0, w_a$ free} \\
\cmidrule(r){2-3} \cmidrule(r){5-14}
(+ S4/Planck T\&P) & \multicolumn{2}{c}{$\bm{ \sigma(\Sigma m_{\nu}) }$ [meV]} &&  \multicolumn{2}{c}{$\bm{ \sigma(w_0) }$} &  \multicolumn{2}{c}{$\bm{ \sigma(w_a) }$} &  \multicolumn{2}{c}{$\bm{ \sigma(w_p) }$} &  \multicolumn{2}{c}{\bf{1/EP}} &  \multicolumn{2}{c}{$\bm{ \sigma(\Sigma m_{\nu}) }$} \\
\midrule
S4Lens &  \multicolumn{2}{c}{69} && \multicolumn{2}{c}{0.25} & \multicolumn{2}{c}{0.95} & \multicolumn{2}{c}{0.14} & \multicolumn{2}{c}{7.5} & \multicolumn{2}{c}{83}  \\ 
\midrule
 & $\bm{k_{\mathrm{max}}=0.2}$ & \bf0.1 && \bf0.2 & \bf0.1 & \bf0.2 & \bf0.1 & \bf0.2 & \bf0.1 & \bf0.2 & \bf0.1 & \bf0.2 & \bf0.1  \\
\cmidrule(r){2-3} \cmidrule(r){5-6} \cmidrule(r){7-8} \cmidrule(r){9-10} \cmidrule(r){11-12} \cmidrule(r){13-14}
S4Lens + LSST Gold & 36 & 50 && 0.12 & 0.18 & 0.37 & 0.50 & 0.050 & 0.085 & 54 & 23 & 55 & 61  \\ 
+ DESI BAO & 22 & 23 && 0.093 & 0.11 & 0.28 & 0.31 & 0.026 & 0.029 & 138 & 111 & 39 & 41  \\ 
\midrule
S4Lens + LSST Optimistic & 31 & 41 && 0.10 & 0.15 & 0.33 & 0.46 & 0.042 & 0.077 & 71 & 28 & 48 & 54 \\ 
+ DESI BAO & 21 & 22 && 0.085 & 0.11 & 0.26 & 0.31 & 0.024 & 0.028 & 159 & 117 & 37 & 40  \\ 
\bottomrule 
\end{tabular}
\caption{Forecasts of the neutrino mass and dark energy constraints, for different experiment configurations and $k_{\mathrm{max}}$ limits. The first two columns assume the dark energy parameters are held fixed, while marginalizing over six $\Lambda$CDM parameters and linear bias amplitudes in tomographic bins, and the rest of the columns include $w_0$ and $w_a$ as free parameters. } 
\label{table2}
\end{table*}

\begin{figure}
	\includegraphics[scale=0.3]{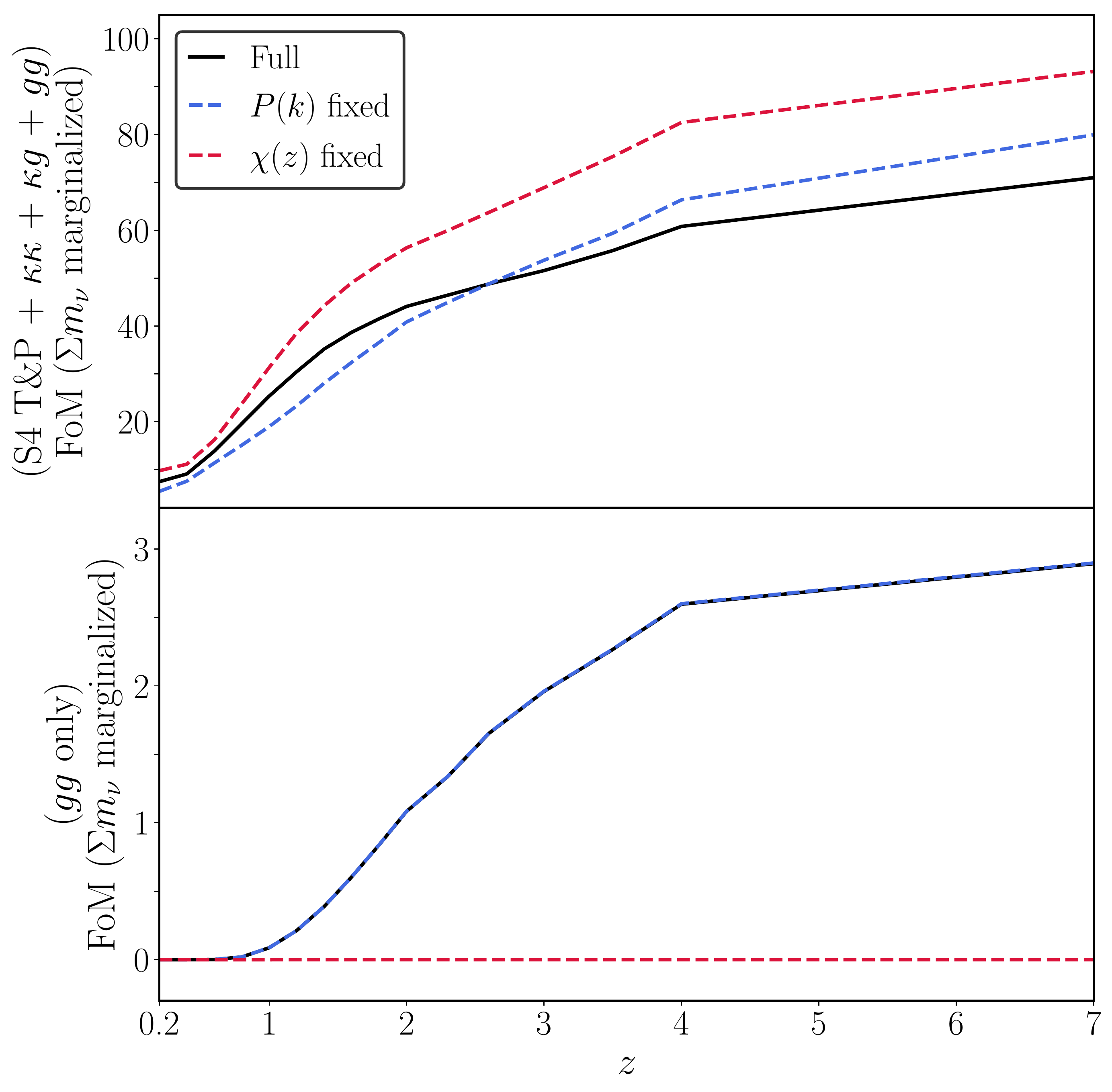}
    \centering
    \caption{ Forecasted DETF Figure of Merit, defined as $[\sigma(w_a) \sigma(w_p)]^{-1}$, with changes to either growth (blue) or geometry (red) disregarded. We also include the results with the "full" Fisher matrix (black), where we apply no restrictions (all effects included), for comparison. \textit{Top}: 1/EP from the combination of S4 primordial CMB, S4 lensing, and LSST clustering (using Optimistic $dN(z)/dz$ with $k_{\mathrm{max}} = 0.2h$Mpc$^{-1}$). We observe the partial cancellation between growth and geometry. \textit{Bottom}: Only with LSST galaxies, ignoring changes to geometry (by fixing $\chi(z)$) makes the constraining power negligibly small. } 
    \label{fig:FoM1}
\end{figure}

\section{Forecast results}\label{sec:forecastResults}


We frame our $w_0$, $w_a$ forecasts in terms of an Error Product (EP), which is inversely proportional to the Dark Energy Task Force (DETF) Figure of Merit (FoM) \citep{albrecht2006report}, defined as the inverse of the area of an error ellipse in the $w_0$, $w_a$ plane. Hence, a higher FoM corresponds to a smaller error.  The EP is a simpler alternative to the FoM, defined as $\sigma(w_a)\times\sigma(w_p)$, where $w_p = w(a_p)$, and $a_p$ is the scale factor at which the uncertainty in $w$ is the least \citep{zhan2006cosmic,abell2009lsst}. In this work, we assume that FoM $\equiv$ 1/EP.

\begin{figure}
	\includegraphics[scale=0.29]{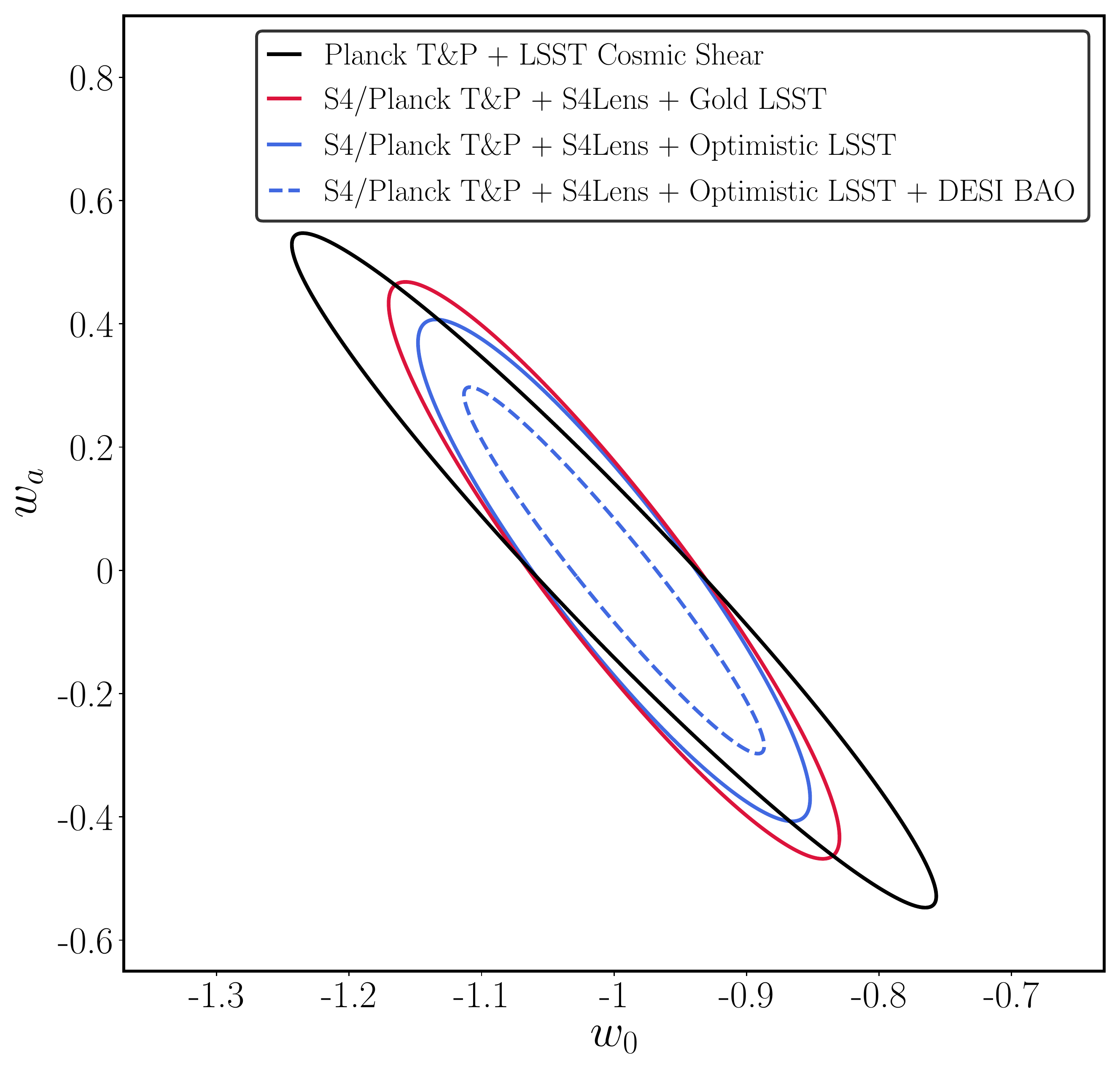}
	\centering
    \caption{$1\sigma$ confidence ellipses in the $w_0 - w_a$ plane, with different survey configurations. We observe that dark energy constraints from LSST clustering in combination with CMB-S4 lensing (blue and red) are comparable to those from the LSST cosmic shear data (black).}
    \label{fig:FoM3}
\end{figure}

Table~\ref{table2} and Figure~\ref{fig:FoM3} present constraints on the the neutrino mass dark energy equation of state, marginalized over $\Lambda$CDM parameters and linear galaxy bias amplitudes in all bins, for different experiment configurations and $k_{\mathrm{max}}$ limits. We find that the relative merit of cross-correlating CMB lensing with galaxy clustering is huge; with $k_{\mathrm{max}}=0.2h$Mpc$^{-1}$, combining the galaxy clustering from the LSST Optimistic sample and CMB-S4 lensing can achieve the FoM (1/EP) of 71. We assume that the primordial CMB information is included in all forecasts.

\subsection{Growth or Geometry?}\label{sec:distanceOrGrowth}

\comment{
\begin{figure}
	\includegraphics[width=\columnwidth]{FoM_of_z_modified_wa1_arrowed}
    \caption{Comparison of FoM created with the ``Full'' $w$ derivatives, and with $w$ derivatives constrained in various ways.  Here we show 1/EP $\propto$  FoM, for the LSST Gold $dN(z)/dz$, using the linear $P(k)$ power spectrum, with $k_{max} = 0.30h$Mpc$^{-1}$.  The gTE and $\kappa$gTE cases have been divided into two panels, upper and lower.  Arrows indicate changes that happen to the FoM in the cases that $w$ derivatives are: 1. Modified so that changes to $w$ only affect the amplitude of $P(k)$, and not its shape, and 2. Also modified so that distances are held fixed. The non-arrowed lines are for cases such that distance is held fixed but $P(k)$ can vary in both shape and amplitude, and for the reverse of that, letting distance vary but keeping $P(k)$ fixed. Note that in the gTE case, these two lines are nearly coincident with the two arrowed lines.}
    \label{fig:FoM1}
\end{figure}
} 

To investigate how sensitive our forecasts are to the distance-redshift relation (geometry) and the amplitude of the matter power spectrum as a function of redshift (growth), we make the following different types of forecasts: ``$P(k)$ fixed'', for which we do not allow the power spectrum to change as $w$ changes, ``$\chi(z)$ fixed'', for which we do not let the distance-redshift relationship change as $w$ changes, and ``Full'' for which we apply no restrictions.


Figure~\ref{fig:FoM1} shows the result of FoM forecasts for all three cases, using the LSST Optimistic $dN(z)/dz$ with $k_{max} = 0.2h$Mpc$^{-1}$. We present FoM values as functions of redshift: at each redshift, only bins at and below that redshift are included, and FoM increases as we extend the redshift lever arm and thereby include more galaxies, as expected.  

The top panel of Figure~\ref{fig:FoM1} shows the forecasts of CMB lensing and LSST clustering combined. We note that both geometry and growth play a significant role, as can be seen by comparing the red and blue curves; the ``Full'' Fisher matrix (black curve), which include all of the effects, appears to have less constraining power than either geometry or growth, suggesting that there is a partial cancellation at play.  A similar cancellation was noted for WL observables in \cite{simpson2005illuminating,matilla2017geometry} and we find that this applies to clustering measurements as well. In short, we note that the partial cancellation between growth and geometry effects that has been noticed before also appears in our S4 lensing + LSST forecasts.

We also find that inclusion of CMB lensing increases the FoM by a factor of 3-4 (over the S4 primary CMB + $gg$ result not shown in Figure~\ref{fig:FoM1}), suggesting that cross-correalations between CMB lensing and galaxy clustering provide a very competitive dark energy probe. The bottom panel shows that with only LSST galaxies, ``$\chi(z)$ fixed'' case has a negligibly small constraining power, as we cannot gain any dark energy information if the distance-redshift relation is fixed, and the amplitude of power spectrum is degenerate with bias.

\begin{figure}
	\includegraphics[width=\columnwidth]{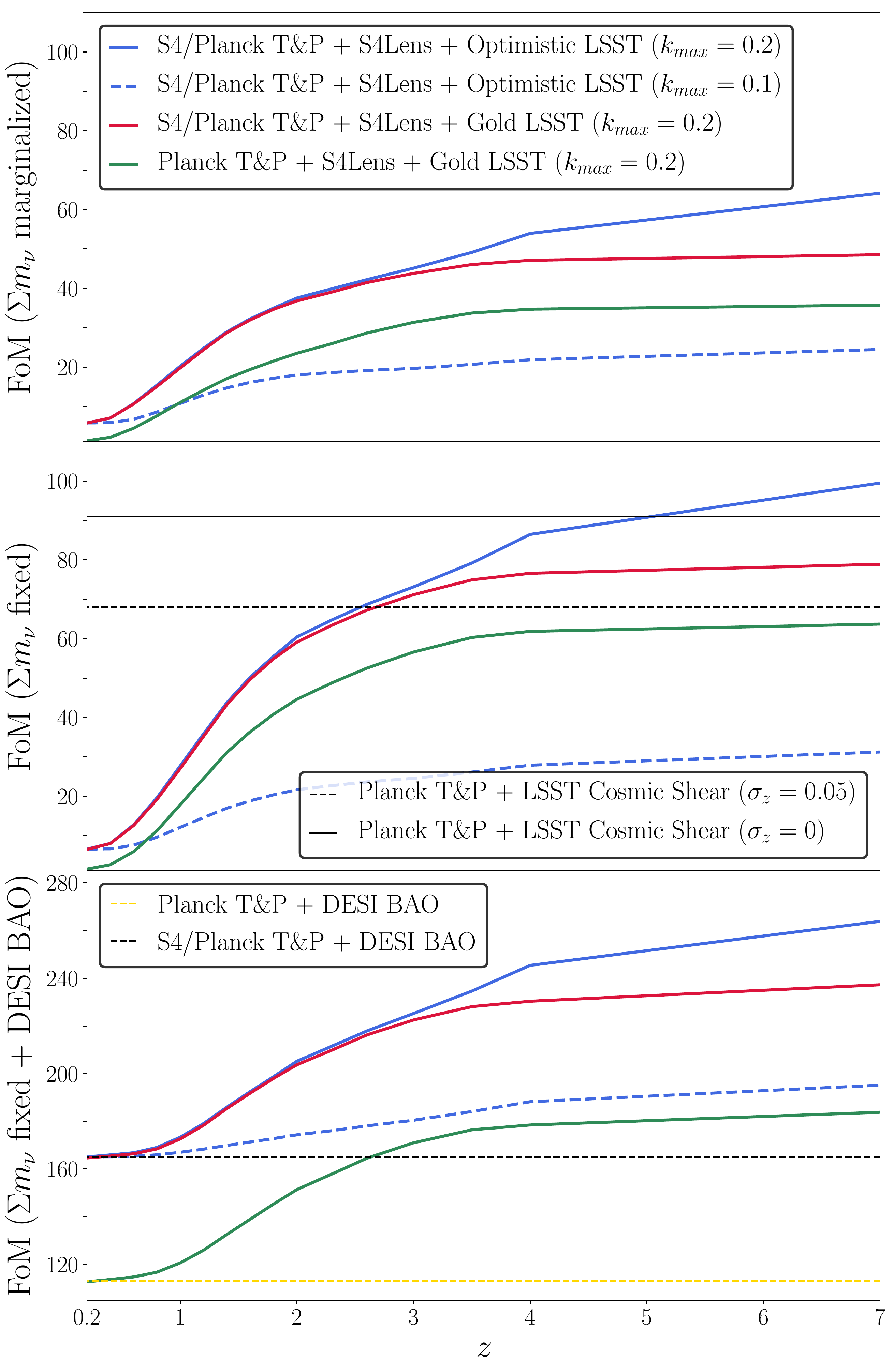}
    \caption{Forecasted DETF Figure of Merit for different experiment configurations and $k_{\mathrm{max}}$ limits. \textit{Top}: Forecasts with the neutrino mass sum marginalized. Addition of galaxy bins at higher redshift extends the redshift lever arm, resulting in a greater constraining power. \textit{Middle}: Forecasts with the neutrino mass sum fixed. Our results are at a similar level to the forecast with LSST weak lensing combined with Planck measurements \citep{zhan2006cosmic} (black). \textit{Bottom}: Forecasts with the DESI BAO measurements included. With the S4 primary CMB data, we gain a noticeable improvement in forecasts relative to the Planck data. }
    \label{fig:FoM2}
\end{figure}

\subsection{Comparisons with Galaxy Weak Lensing Forecasts}\label{sec:comparisonWL}

\cite{zhan2006cosmic} presents LSST Cosmic Shear + \textit{Planck} forecasts of the EP $[\sigma(w_a) \sigma(w_p)]$, for various levels of photometric redshift error, on which the EP depends. Our forecasts assume zero uncertainty in the redshifts of the observed galaxies. Our redshift bin widths are $\Delta z = 0.2$ for the lowest redshift bins up until $z = 2$, then $\Delta z = 0.3-0.5$ out to redshift $z = 4$, and the final bin width of  $\Delta z = 3.0$ from $z=4$ to $z = 7$. To reduce the sensitivity to photo-z errors, we make the bin widths wider than the expected rms scatter in photo-z errors, but we acknowledge that the effects of photo-z errors are not entirely eliminated. We leave an analysis of quantifying such impacts to future work. \cite{zhan2006cosmic} provides the EP as a function of $\sigma_z/(1+z)$, where $\sigma_z$ is the rms photometric redshift error. For simplicity, we compare against the EP corresponding to two specific values of redshift error: EP $\approx 0.011$ for $\sigma_z/(1+z) = 0$ and EP $\approx 0.015$ for $\sigma_z/(1+z) = 0.05$, and these forecasts appear as black horizontal lines in the middle panel of Figure~\ref{fig:FoM2}, labeled as ``Planck T\&P + LSST Cosmic Shear.'' We find that our results are at a similar level to these cosmic shear forecasts. We also note that the middle and bottom panels of Figure~\ref{fig:FoM2} provides the FoM forecasts with fixed $\Sigma m_\nu$ to make a fair comparison to the forecasts in \cite{zhan2006cosmic}.

However, the forecasts in \cite{zhan2006cosmic} assume slightly different choices for the survey characteristics. For the LSST specifications, \cite{zhan2006cosmic} assumes $f_{sky} = 0.48$ and uses a full-survey galaxy number density of $\bar{n} = 50$ galaxies/arcmin$^2$, while we use $f_{sky} = 0.4$ and $\bar{n} = 40$ galaxies/arcmin$^2$. Moreover, \cite{zhan2006cosmic} uses a fiducial bias function of $b(z) = 1 + 0.84z$, whereas we use $1 + z$. The differences in $\bar{n}$, $f_{sky}$, and $b(z)$ should each affect the values of the FoM, but only to a small degree. Changing our forecast parameters to more closely match \cite{zhan2006cosmic}'s would change our forecasted FoM values somewhat.

As an alternative way to compare our forecasts with the LSST cosmic shear forecast, we also present $1\sigma$ error ellipses in the $w_0$-$w_a$ plane.
Figure~\ref{fig:FoM3} includes our forecasts with different survey configurations (assuming $k_{max} = 0.2h$Mpc$^{-1}$), for both the LSST Gold and Optimistic samples. 
Plotted with them is the LSST cosmic shear forecast from \cite{zhan2018cosmology}, and this includes anticipated systematics, such as additive and multiplicative errors in the shear power spectra and uncertainty in the photometric redshift error distribution. This forecast is similar to the one in \cite{zhan2006cosmic}, though the LSST data model is updated, but it does not include galaxy clustering power spectra nor galaxy-galaxy lensing power spectra.  Its error ellipsis appears to have a  similar size as our forecast ellipses, consistent with the result shown in Figure~\ref{fig:FoM2}.


\section{Summary}\label{sec:discussion}

\commentout{
The main focus of this work is to forecast the constraints on the dark energy EoS parameter that can come from the combination of LSST galaxy clustering and CMB-S4 lensing data.  We make several variations on this forecast based on the use of linear or nonlinear matter power spectra, LSST Gold or LSST Optimistic galaxy catalogs, and two different maximum wavenumbers to include in the forecast.  With these FoM forecasts, we compare a fixed-$m_\nu$ version to a prior FoM forecast for LSST weak lensing, and find comparable constraints.  The combination of CMB lensing and LSST galaxy clustering is therefore a complementary method for measuring the dark energy EoS parameter to that of galaxy weak lensing.  However, when we make the comparison of $1\sigma$ error ellipses against an LSST 3x2pt forecast, we see that the LSST 3x2pt + \textit{Planck} forecast has more constraining power than our CMB-S4 + LSST galaxy clustering forecast.

As part of our process for testing the inputs to our forecasts, we vary several parameters, and the resulting relative merit of adding CMB lensing for 5 of the 8 variations that we checked is shown as a function of $z_{max}$ in Figure~\ref{fig:FoM2}. (The others are omitted for clarity.)  To choose the curves to put into this figure, we select the LSST Gold $dN(z)/dz$, nonlinear $P(k)$, $k_{max} = 0.3h$Mpc$^{-1}$ case as our base case (thick solid red), then change the value of one of each of those three variables one at a time to select three more curves, and finally, select one which had all three flipped: LSST Optimistic $dN(z)/dz$, linear $P(k)$, $k_{max} = 0.15h$Mpc$^{-1}$ (thin dash-dot blue). 

One of the parameters that we vary is whether or not we invoke the \textsc{Halofit} nonlinear evolution model when creating matter power spectra.  Of the three parameters, this one appears to have the least impact on its own.  This can be seen in Figure~\ref{fig:FoM2} by comparing the thick solid red and the thin solid red curves.  For the $k_{max} = 0.3h$Mpc$^{-1}$ cases, the linear version has a higher ratio than the nonlinear version (at high redshift), but for $k_{max} = 0.15h$Mpc$^{-1}$, the nonlinear version is higher (comparison not shown).  


The difference between the results produced by using the two galaxy distributions, LSST Gold and LSST Optimistic, is slightly larger than the difference  between results produced by using the two options for linear or nonlinear power spectra, up to about redshift 3.5, where the Optimistic curves can be seen to sharply bend upward.  This difference can be seen in figure~\ref{fig:FoM2} by comparing the thick solid red and the thick dashed blue curves.  The LSST Optimistic $dN(z)/dz$ does better (has a higher ratio) than the LSST Gold in every case, which is expected due to the higher galaxy density and hence higher signal to noise ratio in every redshift bin.  Most notably, the LSST Optimistic version shows improvement in the FoM ratio out past a redshift of 4, whereas the LSST Gold versions are nearly flat after redshift 4.  This increase in the ratio of the Optimistic curves can be attributed to changes in FoM($\kappa$g) only, which can be seen by looking at the previous figure that does not show ratios, Figure~\ref{fig:FoM1}.

The LSST Optimistic curves in Figure~\ref{fig:FoM1} turn sharply upward at redshift 4.  This is actually an artifact of how we create the redshift bins, and how we select the values of $dP(z,k)/dw_0$, $dP(z,k)/dw_a$ at the bin redshift medians for use in our Fisher analysis.  If we instead allow these derivatives to vary smoothly with redshift over the width of each bin, this sharp corner at redshift 4 does not appear.  However, the Optimistic curves do still go to higher FoM than the Gold curves.  This increase in FoM shows that by adding more redshift bins beyond the range at which dark energy is acting, $\kappa$g is able to extract more information about dark energy from the region in which it is acting than is possible without CMB lensing.  We have identified two possible ways that this might be happening.  This may be due to an effective reduction in the noise in the lower-redshift bins which is accomplished by identifying a portion of the CMB lensing signal with matter at higher redshift (like a bin-by-bin de-lensing of the CMB), which increases the FoM, or it could be the result of breaking the degeneracy between the power spectrum amplitude and the galaxy bias.

The difference between results caused by using the two different maximum wavenumbers of $k=0.15h$Mpc$^{-1}$ and $k=0.30h$Mpc$^{-1}$, at the $z=7$ end of Figure~\ref{fig:FoM2}, makes differences in the FoM ratios that are of similar size to those produced by the selection of redshift distribution when comparing the LSST Gold curves, and of slightly larger difference when comparing the LSST Optimistic curves (not shown in plot).  The difference between the different $k_{\mathrm{max}}$ values can be seen in the plot by comparing the thick solid red and the thick dash-dot green curves (or noting that both of the $k_{max} = 0.15h$Mpc$^{-1}$ curves are dash-dotted, whereas the remaining 3 curves have $k_{max} = 0.15h$Mpc$^{-1}$ and are solid or dashed).  Interestingly, the curves with the higher $k_{\mathrm{max}}$ value show smaller improvement from adding CMB lensing than do the curves with the lower $k_{\mathrm{max}}$ value, in every case.  The reason for this is just that the g case with $k_{max} = 0.15 h$Mpc$^{-1}$ by itself does not have a very good FoM (omitted from Figure~\ref{fig:FoM1}).  Since they are so low, when they go into the denominator in the ratio, they boost up the ratio, making it look like the low-$k_{\mathrm{max}}$ case is doing better than the high-$k_{\mathrm{max}}$ case in Figure~\ref{fig:FoM2}, but it's only doing well relative to something that's not that good to begin with.

When we compare the forecasted FoM for the $\kappa$g observables (which contain CMB lensing, redshift-binned galaxy clustering, and the primary CMB) and compare them to the g observables (which have all of those except CMB lensing) we find a relative improvement in FoM by a figure of 2-4, depending on the forecast settings.  There are two main causes for this gain in FoM:  one is that making multiple independent observations causes a simple gain in S/N.  A larger gain comes from the degeneracy breaking that happens between the amplitude of the power spectrum and the galaxy bias.

We also compare our forecasts directly against some that were made with similar assumptions for LSST, which we call LSST WL and LSST WL+BAO (or 3x2pt). We find that our best CMB-S4+LSST galaxy clustering FoM forecasts (with $m_\nu$ fixed) are of the same order of magnitude as those of LSST WL (LSST shear + \textit{Planck}), while the LSST WL+BAO forecasts are significantly better than ours.  One question which we have not addressed is the potential benefit of combining all of these data sets.  In fact, this has already been addressed by others.  For example, \cite{mishra2018neutrino} did a set of forecasts which show the relative strengths of using various sets of observables.  One such comparison can be made in the case where their parameter set included \LCDM, $\Sigma m_\nu$, $w_0$, $w_a$, and curvature ($\Omega_k$).  For $\sigma(w_0)\ (\sigma(w_a))$, they forecast a value of 0.10 (0.26) for LSST+CMB-S4, whereas the value for CMB-S4 alone was 1.14 (2.46), and 0.11 (0.33) for LSST shear + galaxy clustering (without CMB lensing).  

The full combination (LSST+CMB-S4) does the best, beating the LSST shear + galaxy clustering (without CMB lensing) combination in $\sigma(w_0)$ constraints by $10\%$, and $\sigma(w_a)$ by $21\%$.
Unfortunately, it is difficult to compare this forecast directly to ours, due to the different assumptions involved.  What is apparent though, is that the combination of CMB-S4 + LSST shear + LSST galaxy clustering will do better than everything else we have shown here.

Another focus of the work is to investigate by what physical process does our combination of observables detect and measure the value of the EoS parameters $w_0$ and $w_a$.  The EoS parameter affects both the expansion rate of the universe and hence the distance-redshift relation that applies at every epoch of cosmic expansion, as well as the growth of cosmic structure.  We further break down this growth, as it manifests in the matter power spectrum into the growth of the amplitude of the power spectrum, and the change in the shape of the power spectrum due to the difference in growth rate at different scales.  
	
In order to discover by which physical effects the FoM constraints are arising in our forecasts, we create modified forecasts which hold various effects of changing the EoS parameter fixed.  Since the value of $w$ affects the expansion rate of the universe as well as the growth of structure (both in amplitude and shape), we make forecasts where geometry is fixed (no change to fiducial expansion rate) but growth is not, and where growth is fixed (both amplitude and shape) but geometry is not.

The FoM for each of these cases as well as for the full physical case all increase considerably due to the addition of CMB lensing.  We note that the ``$P(k)$ fixed'' version does not directly gain by the breaking of the degeneracy between amplitude and bias, since it is the only variation in which there is no EoS parameter-induced change in amplitude allowed.  However, even though there is no direct gain from breaking this degeneracy, the gain in information about the amplitude allows for gain in information about the other parameters, including the dark energy EoS parameter.

	} 
	
We have studied the prospect for CMB Lensing Tomography to constrain the dark energy parameters by combining LSST redshift-binned galaxy clustering maps and CMB-S4 convergence map.
Although one might expect that the dominant contribution to dark energy constraints would come from the
determination of the matter power spectrum as a function of redshift, the observable statistical properties that we consider (auto- and cross-power spectra) are also sensitive to the distance-redshift relation.  We find that a comparable amount of information about $w(a)$ comes from geometrical features as comes from growth. 

To conduct such analysis, we need to take care not to artificially break degeneracies of galaxy biasing with the amplitude of the matter power spectrum. This danger is present due to the low dimensionality of our parameterization of the bias-redshift relation. In our analysis, we use one parameter for each bin, with a precisely fixed and known redshift dependence within each bin.
To avoid an artificial breaking of degeneracy, we remove variation of the redshift dependence of $P(k,z)$ within a redshift bin. As the dark energy equation of state parameters vary,
we adjust the amplitude of $P(k,z)$ at the center of each redshift bin, while keeping the shape unchanged.  

We find that large angular scales are particularly important for the study of dark energy. Uncertainties associated with non-linear evolution and galaxy biasing on small scales lead us to ignore, in our forecasting, wavenumbers larger than $k$=0.2 $h$/Mpc. Another approach would be to increase the maximum $k$ value as redshift increases, since the scale of non-linearity moves out to higher $k$. However, at higher redshift, the mean bias factor of the galaxies in the catalog increases, reducing tolerance to errors in the modeling of galaxy bias \citep{modi17}. The choice of fixed maximum $k$ means that there is effectively a maximum value of $l$ for each redshift bin, which is an increasing function of redshift. 

Finally, we present the DETF Figure of Merit for different experiment configurations and find that adding CMB lensing information to LSST clustering increases the FoM by roughly a factor of 3-4. 
We also show that our result is comparable to those from LSST tomographic cosmic shear, suggesting that the combination of CMB-S4 lensing and LSST clustering is a competitive probe of dark energy with very different systematics, and therefore highly complementary to the traditional analyses.


\section*{Acknowledgements}


We had very useful conversations with Emanuele Castorina, Enea Di Dio, Emmanuel Schaan, Marcel Schmittfull and J. Anthony Tyson.
SF acknowledges support from the Miller Institute at UC Berkeley, the Berkeley Center for Cosmological Physics and the Physics Division at Lawrence Berkeley National Laboratory while this work was being completed. LK's work was supported in part by NSF award 20010015 and by the U.S. Department of Energy Office of Science.
We acknowledge use of the publicly available \textsc{camb} Python wrapper \cite{cambWeb,pyCambWeb,lewis2000efficient} and \textsc{quicklens} \cite{quicklensWeb,ade2016xv}.

\section*{Data availability}

The data underlying this article will be shared on reasonable request to the corresponding author.




\bibliographystyle{mnras}
\bibliography{crosscor_galaxies}







\bsp	
\label{lastpage}
\end{document}